# Induced eccentricity splitting in disordered optical microspheres for machine learning enabled wavemeter

Ivan Saetchnikov[1], Elina Tcherniavskaia[2], Andreas Ostendorf[3], and Anton Saetchnikov*[3]

[1]*Radio Physics Department, Belarusian State University, 220064 Minsk, Belarus*
[2]*Physics Department, Belarusian State University, 220030 Minsk, Belarus*
[3]*Chair of Applied Laser Technologies, Ruhr University Bochum, 44801 Bochum, Germany*
*Correspondence to: Anton Saetchnikov: anton.saetchnikov@rub.de

## Abstract

Accurate measurement of light wavelength is critical for applications in spectroscopy, optical communication, and semiconductor manufacturing, ensuring precision and consistency of sensing, high-speed data transmission and device production. Emerging reconstructive wavemeters synergize physical systems capable for pseudo-random wavelength dependent pattern formation with computational techniques to offer a promising alternative against established methods such as frequency beating and inteferometry for high-resolution and broadband measurements in compact and cost-effective devices. In this paper, we propose a novel type of compact and affordable reconstructive wavemeter based on the disordered chip with thousands of high quality-factor whispering gallery mode microcavities as physical model and a hybrid machine learning approach utilizing boosting methods and variational autoencoders implemented as wavelength interpreter. We leverage eccentricity mode splitting obtained via controllable deformation of the spherical microresonators in order to ensure the uniqueness of the wavelength patterns up to ultra-wide (∼100 nm) spectral window while guaranteeing high (∼100 fm) intrinsic sensitivity. The latter allocates the proposed model right next to the ultimate reconstructive wavemeters based on integrating spheres, but with superior miniaturization options and chip-scale integrability.

## Introduction

Precise definition of the light wavelength is essential for many scientific and industrial applications and is carried out using dedicated instruments known as wavemeters [1]. In scientific research and analysis, the wavemeters are frequently utilized in spectroscopy measurements, metrology, or as calibration tools to maintain the precision of optical setups [2]. Nowadays, the wavemeters have a crucial role in the actively developing quantum research, where precise control and measurement of the wavelength are essential for experiments related to quantum information, computing, and sensing [3–5]. There are two main configurations for high-accuracy wavemeter construction known: optical heterodyne/beating detection scheme and interferometry [6].

The optical beating method, which measures the beating frequency between tested and reference lasers, offers ∼fm accuracy and high measurement stability. However, this approach has limited measuring range of a few nanometers near the reference wavelength and requires high coherence of the tested laser. The latest advances rely on frequency comb devices as reference sources, extending the measuring range to tens of nanometers and enabling integrated solutions [7]. Unlike combs produced from cavities with a physically limited repetition rate, recently evolved electro-optic frequency combs allow for freely chosen repetition rates driven by a radio frequency (RF) signal and offer ultrafine comb lines for sub-fm resolution [8, 9]. However, this performance can only be achieved when the pump laser power, frequency, and laser-cavity detuning are subjected to rigorous and precise control. The interferometry-based wavemeters are constructed on Michelson, Fabry-Perot, and Fizeau schemes. Among them the Fizeau method, offering a simple setup, broadband detection, and real-time measurement, is the leading interferometry approach for fast laser wavelength detection with sub-10 fm resolution and adopted for commercially available instruments [10]. Interferometric wavelength detection, though, strongly depends on instrument stability, often requiring frequent calibration due to thermal fluctuations. Vacuum chambers and reference sources improve stability, but modern interferometry-based wavemeters remain bulky, expensive, and alignment-sensitive, limiting integration and adoption. This highlights the need for compact, affordable, and reliable designs.

Spectrum analyzers using computational methods to reconstruct light wavelengths from pseudo-random (speckle) patterns enable high-resolution measurements in compact devices with improved dynamic range [11, 12]. Speckle is obtained via interference of multiple reflections/refractions by directing the analyzed light through a disordered medium such as multimode fiber [13–16], photonic crystal cavity array [17, 18], diffuser/scatterer [19, 20] or an integrating sphere [21, 22]. Small wavelength variations cause random changes in speckle patterns, providing strong spectral variability. In such cases, the interpreter linking speckle patterns to wavelength is the key element. Majority of the data-driven methods analyze speckle images across wavelengths to compute the transmission matrix (TM) of the disordered medium [14, 20, 23, 24]. The TM method works effectively assuming uncorrelated speckle patterns for different wavelengths, which fails to hold at small ($\lesssim$ pm) wavelength separations. To overcome this, advanced computational techniques, e.g., machine learning (ML), can be implemented. Different physical principles supplemented with various unsupervised [22, 25] or supervised models [15, 16, 19] have been reported, where the wavelength identification has been considered so far as a classification problem within a limited spectral window, lacking continuous, high-resolution determination across a broad bandwidth in a compact system. The reported reconstructive wavenumbers are extremely diverse in resolving from ~pm down to ultimate values of sub-fm with additional speckle preprocessing under particular constraints. Very recently classical Pearson correlation coefficient has been reported to be a robust measure to determine natural performance of different physical principles of the speckle-based wavemeters [26]. Among diffusers, multimode fibers and integrating spheres, the latter offer the highest intrinsic sensitivity (sub-100 fm), which is one order of magnitude better than for the second-best multimode fibers.

Whispering gallery mode (WGM) microresonator can be employed in each of the mentioned wavemeter techniques and is prospectively capable to harmonize the advantages of different approaches. Inherently, this is an interferometer where the light travels across circular geometry with low energy dissipation and high quality- (Q-) factors [27, 28]. Each WGM microcavity is characterized by the periodic spectral pattern apart by free spectral range (FSR) that depends on the geometrical and optical properties of the microcavity, optical properties of the environment and light coupling conditions that enables high-precision sensing [29–31]. On the one hand, the use of an appropriate material and the maintenance of well-controlled fabrication conditions turns WGM cavity into the main frequency comb source and enables beating method wavemeter [8, 9]. On the other hand, the natural variation between microresonators and random surface inhomogeneities associated with the fabrication limitations, converts the microscale size cavity into a scattering medium with extremely long ($\gtrsim$100 m) optical path and a source for unique spectral pattern. The WGM microcavity can also generate unique temporal speckles through rapid modulation of the laser frequency [32]. Temporal domain representation combined with the ML model incorporating a modified loss function that includes TM-calculated responses, enables resolution down to $\sim$ fm. However, the WGM cavity's pattern is confined to the vicinity of resonance peaks and suffers from spectral gaps and periodicity, limiting its wavelength determination to a narrow range, typically a few pm. Integrating multiple microcavities can produce a pseudo-random, wavelength-dependent pattern [33–35], but the wavelength reconstruction performance of such patterns remains inferior to that of other physical principles.

In this article, we propose a novel physical model based on disordered set of high-Q microresonators with eccentricity modes splitting for realization of the compact, affordable and broadband reconstructive wavemeter with intrinsic sensitivity of ~100 fm. It leverages multiplexed microresonator imaging to generate unique multidimensional wavelength patterns by simultaneously probing thousands of high-Q optical microresonators, previously introduced by us as an environmental sensing instrument [36–39]. We introduced externally induced deformation of spherical microcavities to observe eccentricity mode splitting in the spectra to both improve intrinsic sensitivity by orders of magnitude compared to non-deformed microresonators and to enable broadband detection. The proposed principle is inspired by the frequency combs in optical microresonators, which allows the beating methods to guarantee high wavelength determination precision in broad spectral range. The eccentricity mode splitting phenomenon is known for weakly deformed spherical microcavities (i.e. ellipsoid), where inclined closed circular modes (quasimodes) are turned into open-ended helices winding up on caustic spheroid [40, 41]. This leads to violation of the degeneracy for the modes that propagate in different equatorial planes and resulting in a mode splitting of the azimuthal mode in different equatorial planes. It appears in the spectrum as the side resonance peaks with the period defined by the ellipicity degree. Commonly, mode splitting has been used in characterization of the fabrication quality of microresonators with the goal to avoid the splitting phenomenon as undesirable [42,43]. Hence unlike classical frequency combs, the mode splitting induced comb does not impose strict requirements to enable its observation and does not require any extra pump laser for its excitation. The intrinsic sensitivity puts the proposed model right next to the integrating spheres, which are the most accurate for reconstructive wavemeters, but with improved integrability. The unique combination of the disordered microresonator chip with eccentricity mode splitting as a source for unique wavelength-dependent responses, together with specifically designed interpreter that includes transition to the latent space, variational autoencoders dedicated to guarantee consistent and extensive calibration dataset, and the regression model based on boosting methods provides an avenue for high precision wavemeter in the ultra-wide spectral regions up to $\sim$100 nm.

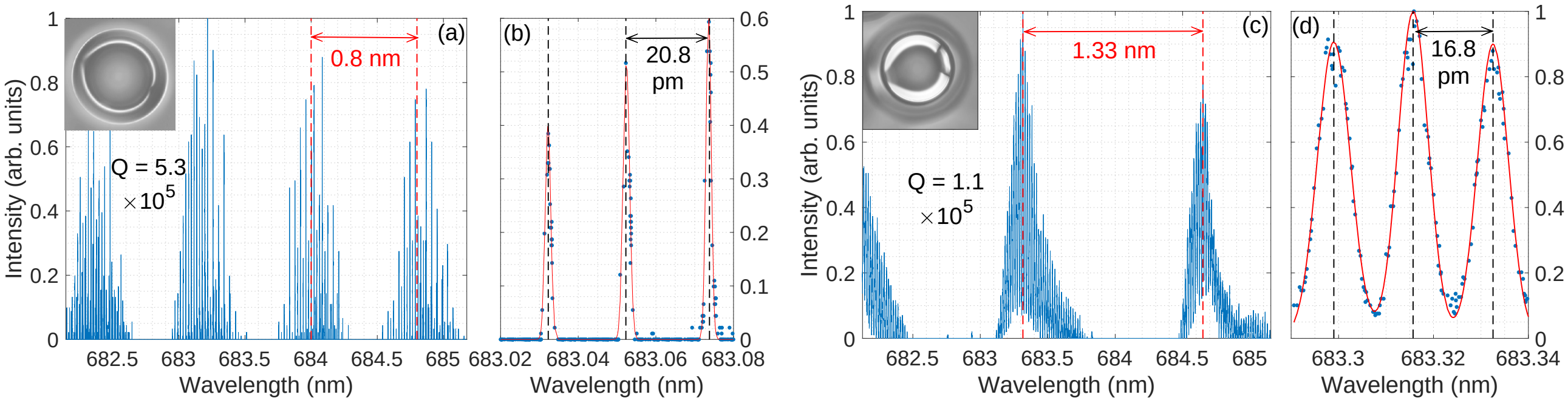

**Fig. 1.** WGM spectra with eccentricity mode splitting for two exemplary glass microspheres of different dimensions. The spectra over several free spectral ranges along with zoom into the splitting region for bigger and smaller microcavities (a), (b) and (c), (d) correspondingly. Blue lines and dots represent experimental results, red lines - fitting of the resonances with Voigt profile. Insets provide the microscope picture of the microresonators highlighting their differences in size.

## Results

### Eccentricity splitting in microspheres

The feasibility for observation of mode splitting for the multiplexed microresonator chips has been primarily studied on example of glass microspheres. Mode splitting for the microcavities on the multiplexed chips has been determined as rare phenomenon where only a few microresonators with clearly detectable mode splitting were identified for each chip. The spectra of two exemplary glass microspheres with mode splitting are represented in Fig. 1.

Both spectra represented in Fig. 1 are characterized by the presence of plurality of the splitted modes of different mean loaded Q-factors ($5.3 \times 10^5$ and $1.1 \times 10^5$) that exist within the splitting windows of comparable width. Difference in size of the microcavities that manifest as different azimuthal free spectral range ($FSR1$) leads to different frequency of splitting blocks occurrence within the same spectral range. Here splitting block is shaped by the loaded Q-factor and the spacing between the splitted modes ($FSR2$). Similar pitch value for splitted modes (20.8 and 16.8 pm) in case of lower Q-factor ($1.1 \times 10^5$) results in partial overlapping of the resonance lines and formation of the broad resonance profile Fig. 1 (d). It is still characterized by high-frequency intensity modulation due to mode splitting, which decreases in severity with increase of the spectral overlapping of the splitted modes. As result, frequent splitting that appears for slightly deformed cavities and insufficiently high Q-factor can lead to the transformation of the splitted spectrum into low Q-factor resonance profile. Thus, efficient generation of splitted spectrum is possible when the optimal ratio between Q-factor and deformation of the microresonator is ensured.

As a result of sphere deformation, the ellipsoid surface with different semi-major and semi-minor axes is obtained. Thus, the orientation of the excitation beam plane with respect to these axes will define the effective ellipticity degree. To demonstrate the reliance of the ellipticity degree and mode splitting mechanics in the WGM spectrum, we rotated the chip relative to laser propagation plane. Thereby different ellipticity levels along the WGM propagation path for the distorted microspheres on the chip can be realized. Variations of the mode splitted specta for a representative deformed microcavity for four different orientations of the chip relative to the beam are represented in Fig. 2.

The results show clear variations in the mode splitting when the excitation plane is varied. In particular, by increasing the rotation angle (entirely defined by selection of the initial) the limited window for mode splitting expands and eventually takes up the entire FSR of the WGM. In the demonstrated example the $FSR2$ changes from 8 to 65 pm when different orthogonal planes in the microresonator are utilized to excite the WGM. Under consideration, that $FSR1$ for the depicted microcavity accounts $\approx 0.8$ nm, the effective eccentricity level changes from 1% to 8%. It has been revealed that the splitting step coincides for modes of different orders. Position of the neighboring splitted modes of different orders are marked as vertical lines of different colors in Fig. 2. Severe deformations of the selected microsphere that manifests with up to 8% eccentricity level explains the non-uniformity of the splitting step variations with changes of the microcavity orientation. In particular, by changing the orientation of the microsphere on 24° (from 66° to 90°) near the projection with maximum distortion the same 65 pm splitting step was revealed. Comparable rotation of microresonator on 20° (from 0° to 20°) results in more than doubling of the splitting step. Besides the splitting step, the orientation of the microresonator affects the loaded Q-factor where for particular orientation it reaches $1 \times 10^6$. It is, however, weakly related to the eccentricity level and is rather determined by the scattering losses in the plane of the mode propagation.

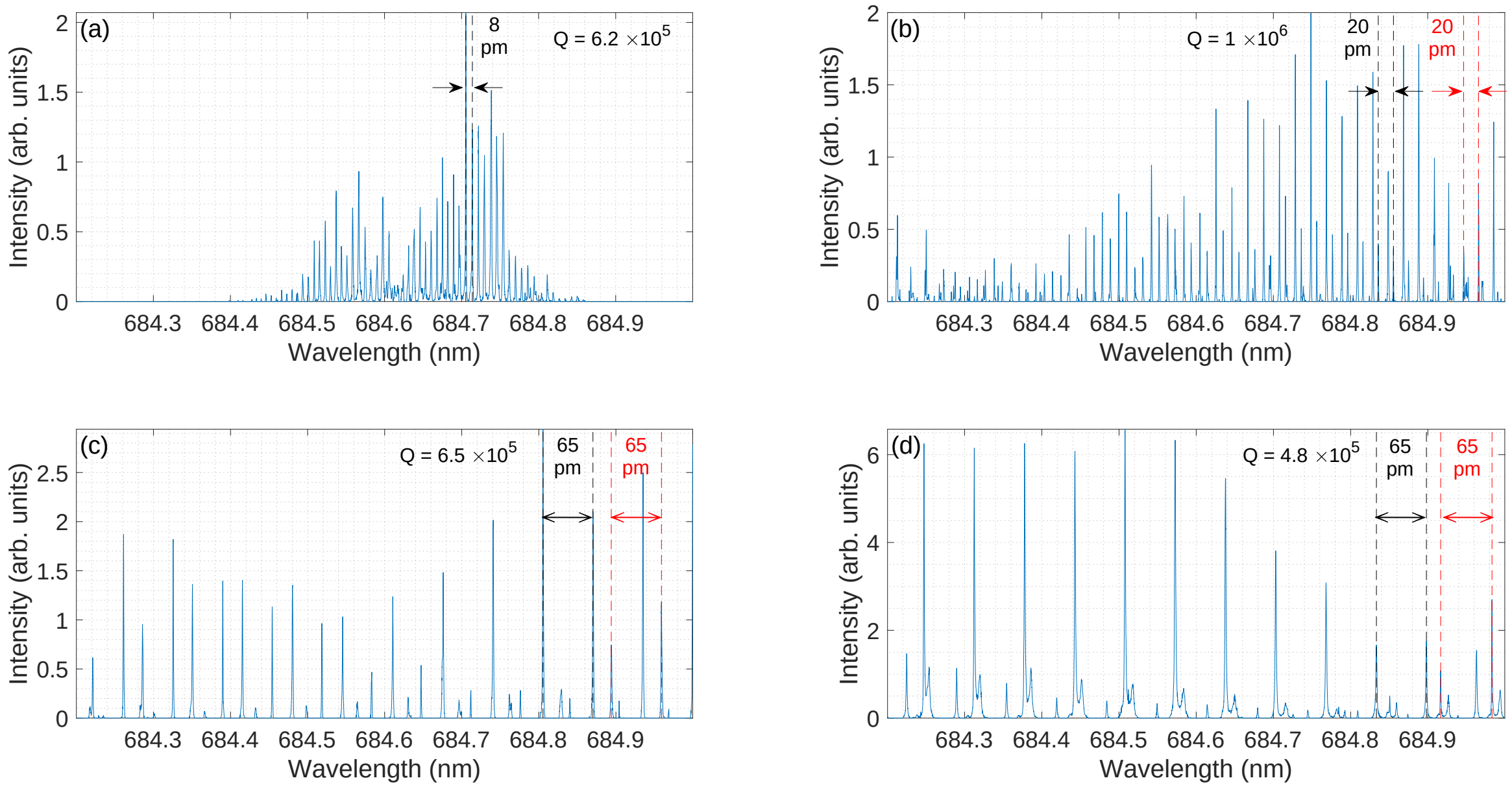

**Fig. 2.** Mode spitted spectra in a deformed glass microsphere at different chip orientations relative to the excitation beam plane. Orthogonal excitation planes in the exemplary glass optical microresonator 0° (a) and 90° (d) along with two states between them 20° (b) and 66° (c) are represented. Black and red colored vertical lines indicate modes of different orders.

### Induced eccentricity splitting

Despite considerable fraction of non-spherical microresonators ($\lesssim$15%) specified by manufacturer, the number of glass microresonators demonstrating spectrally-resolved mode splitting does not exceed 1% on average over the analyzed samples. A comparable ratio of splitted responses to the total number of cavities was determined also for the chips containing PMMA microspheres. This appears due to the low eccentricity level, which leads to partial overlapping of the splitted modes, resulting in broadening of the resonance and limited spectral discrimination of them. Besides this, the purely random nature of the eccentricity splitting on fabricated chips does not allow to find optimum mode splitting performance. Thus, it remains impossible to analyze the effect of spectral mode splitting on the accuracy of the wavelength determination by utilizing non-processed microresonators due to the inability to generate a sufficient set of features.

Unlike the glass with superior mechanical, thermal and chemical stability, the PMMA material can be relatively easily manipulated, e.g. by heating them above glass transition temperature. PMMA begins to soften at the temperatures above 100°C, where at higher temperatures it can begin to melt and deform. Therefore, modified PMMA-based chips with induced eccentricity splitting can be utilized to construct sufficient fraction of the mode splitted spectra in the experimental dataset. In order to achieve an elliptical shape from the spherical one, the particles have to be stretched in one direction typically by applying force in one direction. Thanks to the permanent bonding of the microresonators with the substrate via fixation layer and necessity for subtle deformations only, sufficient eccentricity can be ensured simply by placing the chips with raw PMMA spheres on the hot plate without applying external force. We tested different processing temperatures (from 80 to 250°C) for heating the samples over 10 min in regard with observation of the mode splitting. Thermal treatment at 220°C leads to visually apparent deformation of the microresonator's geometry and consequently high radiation losses. By increasing the temperature to 250°C the microresonators melt down completely. Comparison of the spectral properties of WGM cavity before (0°C) and after thermal treatment with varying temperature from 80 to 180°C is represented in Fig. 3.

Even though the original WGM spectrum contains several modes within the single FSR, the following analysis of the impact of thermal treatment will be focused on the most pronounced among them. After processing the chip at 80°C, the WGM became broader (Q-factor reduces from $1.3 \times 10^5$ down to $1.1 \times 10^5$) and its intensity rises. When increasing the temperature further up to 120°C, the side peaks begin to manifest in the spectrum, indicating splitting, but they still remain poorly resolved. At this point, the intensity of the resonance reaches a maximum that is $\approx 5$ times higher than the resonance signal of the untreated chip with PMMA cavities. At the next state at 140°C, splitted modes drift further apart as a result of higher eccentricity level that shows up in more distinct splitting pattern with the pitch of 7.3 pm ($\approx$ 0.5% eccentricity level), but the splitted modes still overlap via the side tails. At the same time, this state is characterized by reduction of the mode peak intensities to the original level and doubling the Q-factor compared to the original state. The spectrum of the microresonator processed at 160°C is characterized by the similar level of the peak intensities as before. However, the Q-factor rises further up to $4.2 \times 10^5$ and the splitted modes became clearly resolvable without overlaps with $FSR2 = 10.1\ pm$ ($\approx$ 0.7% eccentricity level). Subsequent thermally induced deformations lead to anticipated widening of the spectral pitch between the splitted modes up to 12.2 pm. The prominence of the modes here reduces already by an order of magnitude that could indicate changes in coupling conditions due to cavity deformation. By trying to deform the resonator further, the resonance signal vanishes. Other high-order modes detected in the FSR also undergo changes during thermal treatment, that, however, can be inconsistent with the most prominent mode. In general this leads to observation of the both spectrally-resolved and broadened mode splitting combs within a single FSR. Accounting the results for other PMMA microcavities among different chips, the temperature window from 140°C to 160°C has been defined as optimal to observe spectrally resolved spitting of the modes. Dispersion in the temperature is conditioned by the natural variations of the microresonators. It is important to point out that the Q-factors of the splitted modes after thermal treatment of the PMMA chip exceed the original state. This is caused by the broadening of the peak in the initial state due to the overlap of splitted modes arising from the initial small non-sphericity of the microresonator. Summarizing the results over examined PMMA chips, thermal processing enables to expand the portion of the microresonators with splitted spectra from original $\approx 1\%$ up to $\approx 45\%$, that allows to construct adequate experimnetal dataset.

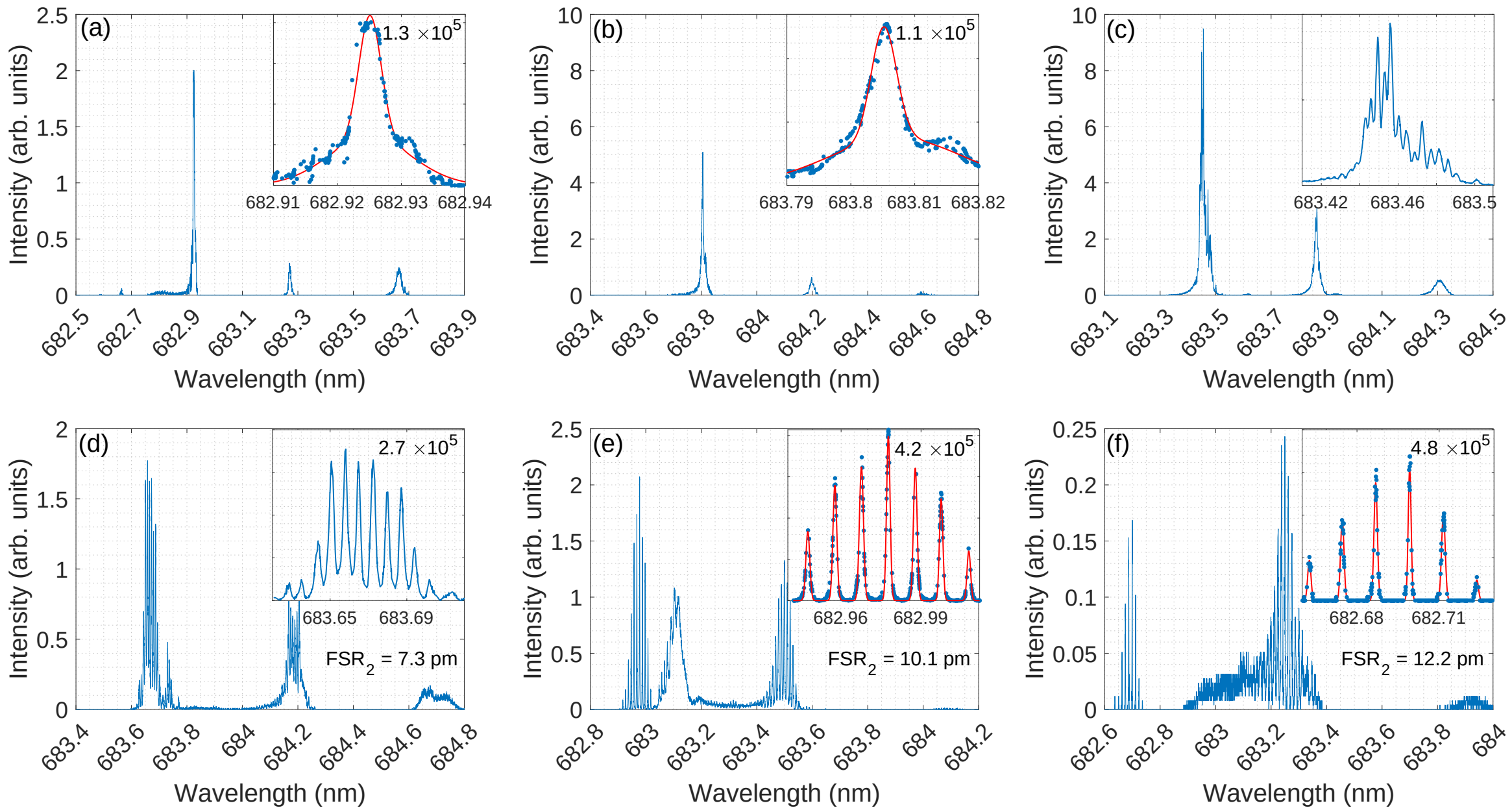

**Fig. 3.** Eccentricity mode splitting for exemplary PMMA microresonator in the initial state (0°C) (a) and heated at different temperatures: 80°C (b), 120°C (c), 140°C (d), 160°C (e), 180°C (f). Blue lines and dots represent experimental results, red lines - fitting of the resonances with Voigt profile and the insets focus on the most prominent mode.

## Models

### Model of the eccentricity splitted WGM spectra

To study the impact of the eccentricity-induced mode splitting on the degenerated WGM spectrum for high-precision wavemeters, a dataset with well-controlled conditions for eccentricity splitting is necessary. While rigorous mathematical models for WGM resonances in optical microresonators are well established and can be found elsewhere, the multiplexed microresonator imaging configuration and eccentricity-induced mode splitting require adaptation to the original WGM model. The Lorentzian broadening mechanism characterizes the WGM spectrum of ideal microresonator with degenerated modes. Natural shape alterations, material inhomogeneity, roughness on the microresonator surface and glass substrate lead to additional inhomogeneous (Gaussian) broadening [44]. The convolution of the Gaussian and Lorentzian profiles is known in spectroscopy as Voigt profile and can be used to describe WGM spectral patterns [45]. This can be simplified to the pseudo-Voigt form, which describes power spectrum of the microcavities $P\omega$ as follows:

$$P = {}_{i=1}^{n} P_0^i \big[\eta_i \frac{\gamma_i 2^2}{\omega - \omega_0^{i\,2}\; \gamma_i 2^2} \\ 1 - \eta_i \exp \frac{-\omega - \omega_0^{i\,2}}{\gamma_g^{i\,2}}\big], \tag{1}$$

where $n$ is the number of modes, $P_0^i$ is the coupled light power of the $i^{\text{th}}$ mode, $\eta_i$ characterizes the ratio between the homogeneous and inhomogeneous broadening mechanisms for the $i^{\text{th}}$ mode, $\omega_0^i$ is the central frequency of the $i^{\text{th}}$ mode, $\gamma_i$ is the full width at half maximum (FWHM) of the $i^{\text{th}}$ mode, and $\gamma_g^i = \gamma_i 2\sqrt{2\ln 2}$.

Initially, the degenerated spectrum with one (main) mode allocated within the cavity's free spectral range ($FSR1$) is simulated. Its spectral position is influenced by a range of independent factors, including resonator parameters (such as size, shape, and material), light coupling settings, and illumination polarization. Therefore, the distribution of the resonance position for degenerated mode ($\omega_0^m$) among the cavities is assumed to be uniform within the $FSR1$. The latter parameter varies for the cavities due to their feature differences and is assumed to follow uniform distribution in the range [$FSR1^{\min}$, $FSR1^{\max}$]. Similarly, the Q-factors of the modes are also anticipated to be uniformly distributed between the minimum ($Q_{\text{tot}}^{\min}$) and maximum ($Q_{\text{tot}}^{\max}$) observable values. The coupled light power for the main mode ($P_0^m$) is also defined as uniformly distributed number from $P_0^{\min}$ to $P_0^{\max}$. When the simulation region exceeds the $FSR1$, main modes with different azimuthal numbers are simulated by allocating them equidistantly on $FSR1$ around $\omega_0^m$.

When the azimuthal mode degeneracy is lifted, the mode splitting shows up. In order to simulate this, we have introduced two parameters to the model. Eccentricity level ($e = \frac{a-b}{a}$), where $a$ and $b$ are the semi-axes, describes the nonsphericity of the microresonator. This parameter is used to define free spectral range of the the mode splitting as ($FSR2 = eFSR1$) [41, 42]. Another parameter is the window for mode splitting ($w$) that indicates the spectral region for phase matched energy transfer from the prism to the microresonators. Within the window $w$ a set of the equidistantly ($FSR2$) allocated central frequencies $\omega_0^i$ around $\omega_0^m$ has been defined. Under the assumption of small deviations from the spherical shape, the resonance line widths of the side modes ($\omega_0^i$) are not expected to deviate considerably from the main mode ($\omega_0^m$) and, therefore, are set to be identical to the main mode. The same is done for ratios between homogeneous and inhomogeneous broadening ($\eta_i$) where all of them are set equal to $\eta_m$. Coupled light power for the modes within the splitting window ($P_0^i$) are expected to follow normal distribution (phase matching conditions degrade with moving away from the central mode) with the maximum at previously defined $P_m^i$. In order to ensure variability among different microresonators, the set $P_0^i$ is defined as the histogram distribution obtained for the randomly generated values. The procedure repeats for other main modes with different azimuthal numbers to obtain the spectrum exceeding $FSR1$. As result, a group of the following parameters $\omega_0^i$ , $\gamma_i$, $P_0^i$, $\eta_i$ is established and noise-free WGM spectrum with lifted degeneracy can be simulated using Eq. 1.

Real spectrum is accompanied by various noise components of both fundamental and technical nature. The impact of the fundamental noise sources depends on the integration time, where noise due to temperature variations becomes more prominent than the shot noise for time scales longer than a few milliseconds. This aligns with the minimum possible averaging time defined by the maximum camera frame rate of the multiplexed microresonator imaging scheme. Technical noise sources are linked to the laser and detector (camera). Laser instabilities, such as wavelength sweeping region repeatability and output power fluctuations, are considered negligible. The laser operates within a spectral range with no notable power fluctuations, and collected spectra are normalized to suppress long-term power variations. Camera-generated noise, described as Gaussian noise, is dominated by read-out noise in the selected operation configuration. As a result, thermoelastic, thermorefractive, and detector noise contributions must be accounted for in the eccentricity splitted WGM spectrum model.

The instability of the resonance central line $\sigma\omega_0^i$ induced by the temperature fluctuations $\sigma T$ through the thermorefractive ($\alpha_n = \frac{dn_{eff}}{dT}$) and thermoelastic effects ($\alpha_R = \frac{1}{R}\frac{dR}{dT}$) is described by equation:

$$\frac{\sigma\omega_0^i}{\sigma T} = -\frac{1}{n_{eff}}\alpha_n\omega_0^i - \alpha_R\omega_0^i. \tag{2}$$

For each frequency $\omega$ in the spectrum that is taken frame-wise by the camera while the laser is being swept, the positions of the mode central frequencies are represented as $\omega_0^i\ \sigma\omega_0^i$. Despite the discrepancy in the effective refractive index $n_{eff}$ among the modes, the modes are assumed to be entirely confined within the cavity, making $n_{eff}$ equivalent to the resonator refractive index $n_r$. Short-term temperature fluctuations have been studied experimentally and determined to follow a normal distribution with a standard deviation of 0.01 K. The impact of thermal fluctuations on the WGM linewidth was analyzed for scattering $Q_{scat}$ and coupling ($Q_{coup}$) limited Q-factors. The radiation $Q_{rad}$ and absorption $Q_{abs}$ limited Q-factors attain $10^{15}$ and $10^{7}$, respectively, which exceed the experimentally observed values for the loaded Q-factors (up to $10^{6}$). By using the expressions for $Q_{scat}$ and $Q_{coup}$ proposed in [46,47] the Q-factor instabilities are calculated to not exceed $\sim 10^{0}$ for the mentioned above level of temperature variations (0.01 K). For this reason, the impact of the temperature fluctuations on the linewidth of the WGMs is omitted in the model. The Gaussian noise representing the camera's read-out noise is generated with the mean value equal to zero. The standard deviation of the noise is estimated from the experimental data by examining signal variations under non-resonant conditions and equals to $\approx 0.001$.

**Regression interpreter model**

Wavelength prediction was treated as a regression task on tabular data, that can be solved by decision-tree-based (DTb) and deep learning approaches (dNN, deep neural network). The detector noise has a vertical projection into the spectrum, whereas the thermal noise has a horizontal one. This presents an additional challenge to data preprocessing, which, in turn, makes dNN less suitable for wavelength prediction. Furthermore, since the features (i.e., microcavities in this case) are essentially equivalent, selecting the most prominent ones for the training dataset results in the lack of generalization due to unique resonator spectra. Given that dNN are more susceptible to inefficient features than DTb models, the latter are deemed to be the optimal baseline approach.

Ensemble boosting methods like LightBoost, XGBoost, and CatBoost are highly accurate and robust, especially for tabular data, and are categorized as DTb methods. Gradient boosting (GB) models differ from other ensemble methods like bagging and stacking by building decision trees sequentially, where each tree corrects errors from previous ones. With well-tuned parameters, this iterative process enables GB models to capture complex patterns efficiently. LightBoost represents a modification of GB that integrates the ”Gradient-based One Side Sampling” and ”Exclusive Feature Bundling modules”. The first module focuses on training examples that yield larger gradients, thereby accelerating training and reducing the computational complexity. The second module employs a boosting approach that combines sparse (mostly zero) mutually exclusive features by bagging these attributes. This enables smart feature selection, which is particularly useful for wavelength prediction, where preliminary feature selection is restricted. These modules make LightGBM both accurate and fast, with improved memory efficiency and scalability over the slower XGBoost.CatBoost can attain enhanced regression accuracy bit for datasets with categorical features.

The tuning process was dedicated to determine the optimal number of boosting iterations with maximum set to 1000. Learning rate, number of leaves, maximum depth of the decision tree, and L1 and L2 regularization parameters were optimized based on the results for a 5-fold cross-validation. It ensures comprehensive coverage of all available wavelengths and thus enhances the tuning relevance. The tree is constructed until it either reaches the maximum depth or is unable to identify an optimal split. By establishing this parameter, the model is prevented from fitting noise in the training data with excessive precision, thereby reducing the probability of overfitting. In order to enhance the robustness of the model, the number of leaves is set to a value that is considerably below power of two raised to the maximum depth. By increasing the maximum depth and the leaves number, the model performance can be improved by the cost of higher probability for overfitting and time required for training. To achieve equilibrium between these elevated parameter values, the minimum data-in-leaf parameter, which establishes the minimum number of observations required in a leaf node, is introduced. Low value of this parameter may result in overfitting, as excessively specific leaf nodes may capture noise and thereby reduce generalization. Conversely, an excessively high value may result in underfitting. To further mitigate the risk of overfitting, regularization was employed using L1 and L2 parameters. The training process with the predefined learning rate continues until either the maximum number of boosting iterations was reached or the performance on the validation test does not improve during the last 50 iterations. Lightboost regression model is built and trained using the following hardware configurations: 14 cores Intel Xeon W-2275 processor, 128 GB of RAM coupled with Nvidia T4 GPU with 16 GB of RAM as well as Nvidia Quadro RTX 6000 with 24 GB of RAM for training and optimizing models. The processing was conducted using the Python programming language, which leveraged the CUDA and PyTorch libraries, along with a set of additional Python libraries, including TensorFlow and Keras.

**VAE model for pattern conditioning**

Large volume of high-quality data is essential for achieving robust and accurate predictions. This is the common challenge in many data-driven inverse problems since collection for such data is a major time-consuming constraint. Also, ensuring the data quality is particularly difficult, as variations in physical constants and experimental conditions can introduce errors and biases, which, in turn, negatively affect the performance of the predictive model. There are two competing approaches to address this challenge. The first involves utilization of the analytical model to generate sufficient data, pre-training the predictive model on this simulated dataset, and then fine-tuning it using a limited set of experimental results. However, given the enormous spectral variability among WGM microcavities, including the presence of multiple modes of various orders and mode splitting, the generated multiplexed spectral data may suffer from assumptions in the WGM spectrum model. A variational autoencoder (VAE)-based architecture has been put forth as a potential solution for data augmentation in the context of limited experimental spectra. The VAE core is two mirrored dNNs referred to as encoder and decoder. The first one transforms input data into a lower-dimensional latent space, the second reconstructs the original data from the latent space. In order to achieve a one-step data augmentation, we propose the inclusion of one-dimensional convolutional layers for both the encoder and the decoder. An alternative to the convolutional is the use of long short-term memory (LSTM) layers, which can be effectively employed for limited spectral width, thereby avoiding the risk of losing long-term information when processing long spectra. This necessitates sampling of the spectral windows from the original sequence, their subsequent training and merging the reconstructed windows to obtain the full spectrum.

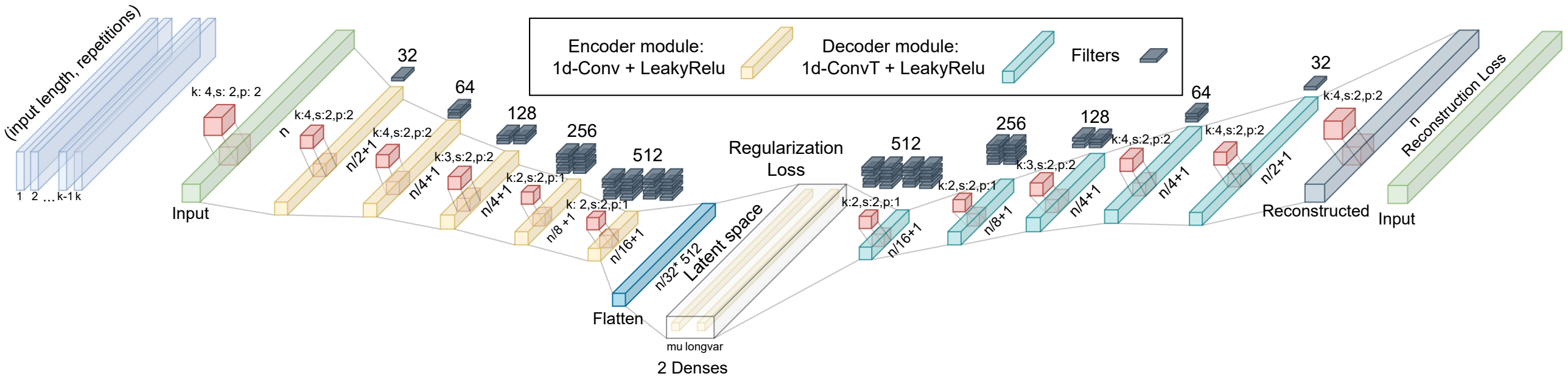

**Fig. 4.** Architecture of the variational autoencoder for data augmentation of the multiplexed WGM spectral data.

Encoder comprises five modules each out of a convolutional layer followed by a LeakyRelu activation function (Fig. 4). The number of filters is increased by a factor of two at each successive layer, with filter sizes of 32, 64, 128, 256, and 512. The sizes of the kernel, stride, and padding were optimized in order to achieve a continuous halving of the input size sequence at each step, reducing it from n to n/32. During the compression process on the convolutional layers, the size of the convolutional kernel has been reduced in order to facilitate the manipulation of local patterns and the extraction of fine-grained patterns in the data. Convolution modules are followed by two fully connected layers in a latent space with a dimension of 2000. The decoder consists of five up-sampled modules of transposed convolutional layers with LeakyReLU activations, featuring filter sizes of 512, 256, 128, 64, and 32. The parameters for kernel size, padding, and stride have been configured in accordance with that defined before for encoder. In particular, the first convolutional layer in the encoder employs the identical parameters as the final deconvolutional layer in the decoder, etc. The VAE is trained using the Adam optimizer with an adaptive learning rate of $1 \times 10^{-3}$ decreasing each 50 epochs on gamma 0.1. The objective is to minimize a loss function comprising two components: the reconstruction loss, based on the cumulative mean squared error (MSE) across all batch sequences, and the regularization loss, derived from the Kullback–Leibler divergence (KLD). The parameter beta, which adjusts the contribution of the KLD term in the total loss, is set to 0.005. The training process of the VAE continues for 150 epochs, continuously updating the trained model with weights that demonstrate the best performance on the testing set until the total training loss reaches a plateau. VAE model is optimized using the same hardware configuration as utilized for the regression model.

## Performance assessment

The evaluation of the performance of the wavelength reconstruction model has been initially performed with a simulated dataset. To achieve this, individual datasets were constructed, each containing spectra of microspheres characterized by certain level of eccentricity ($e$). It has been varied in the range $[5 \times 10^{-4},\ 6.4 \times 10^{-2}]$ with an additional case of $e = 1 \times 10^{-6}$ included to generate the spectral responses of microresonators exhibiting nearly ideal spherical symmetry. Each of these datasets contains WGM responses of 200 microresonators with pre-defined positions of the main modes with Q-factors selected from $[2 \times 10^5,\ 2 \times 10^6]$ and allocated with $FSR1 = 0.8; 1.2$. The coupled light power for the main mode ($P_0^m$) has been searched in [0.125; 1] and $\eta_m$ has been fixed at 0.8. To maintain a reasonable simulation time, the spectral width of the simulated data was limited to 1.2 nm, while the spectral resolution was set at 0.3 pm. Each WGM spectrum has been simulated 2352 times to consider the impact of the spectral noise. The thermoelastic and thermorefractive coefficients were set at $1 \times 10^{-4}\ K^{-1}$ and $-2 \times 10^{-4}\ K^{-1}$, respectively, to accurately model the behavior of PPMA microresonators. The spectral window for the splitted modes $w$ are selected from the range [0.2, 0.7], where the number of the modes within window $w$ was limited to 40 for computational reasons.

The performance of the wavemeter is defined by intrinsic sensitivity of the disordered multiplexed microresonator patterns. This parameter allows to benchmark the proposed physical principle among the other wavemeter design [26]. The results for the similarity (Pearson correlation coefficient) of the multiplexed microresonator intensity patterns over the wavelength change for different $e$ values are represented in Fig. 5. Each curve has been obtained as median value for different positions of the wavelength within the whole simulation region among all sensing repetitions. The results show that the shape of the correlation curve changes significantly when altering the eccentricity level. The system with 200 non-deformed microresonators is characterized by the natural sensitivity of 2.2 pm (defined as half width at half maximum, [17]) and comparable to the mulimode fiber-based speckle wavemeters. By applying minor deformations to the microspheres ($e$ = 0.0005) the correlation curve shows sensitivity worsening. When increasing eccentricity level to 0.001 the correlation curve shows oscillations due to closely spaced resonance peaks. By deforming microspheres further sensitivity improves significantly, where the best intrinsic performance for 200 microresonators has been determined for $e$ = 0.012 at $\approx$ 360 fm level. This is already approaching the accuracy of the integrating spheres which represent the most precise physical principle for speckle wavemeters currently available. At the same time, the splitting peaks regularity may result in multiple crossings of the 0.5 similarity level when a broader range of changes in the wavelength is observed. This is clearly seen for small eccentricity level ($e$ = 0.004) where the correlation value starts to rise towards 0.5 after 2 pm spectral shift. Moreover, the similarity curve does not reflect the perfect similarity of the blank responses that can be observed for $e = 1 \times 10^{-6}$ in the case of the limited amount of microresonators. Thus, the probability for correlation to exceed 0.5 level within the broad range is expected to drop with microresonators number in the system. This is studied with the regression model applied for wavelength reconstruction within the whole range of simulations.

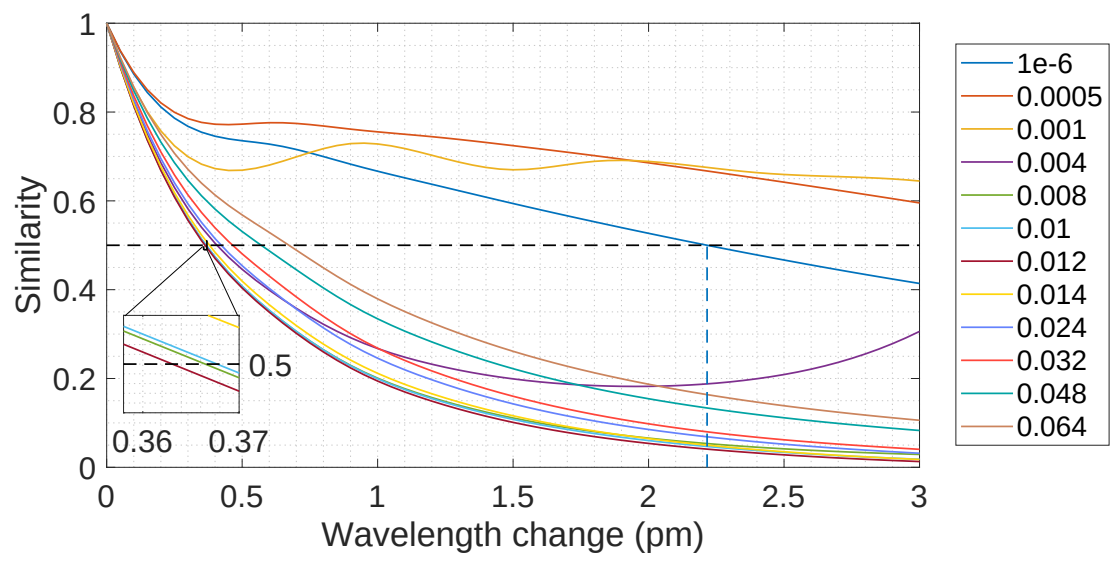

**Fig. 5.** Similarity as function of the wavelength change for the disordeted multiplexed microresonators system with 200 cavities and different splitting levels ($e \in 1 \times 10^{-6}, 6.4 \times 10^{-2}$).

We established an architecture of the regression model common for all simulated datasets. This approach is predicated on the assumption that the highest possible accuracy for each particular set cannot be achieved with the lightest possible architecture. It allows for a comparative analysis of the effectiveness of the regression model for different eccentricity levels. Simulated dataset without mode splitting ($e = 1 \times 10^{-6}$) has been selected to tune LightGBM architecture. This dataset is expected to demand the most complex architecture due to the excess of the spectrally uncertain regions. At the same time, this dataset is lightweight and enables moderate tuning duration. The optimal values for maximum depth and number of leaves have been determined as 30 and 3584, respectively. The other parameters were set as following: minimum data-in-leaf set to 20, learning rate = 0.01, L1 = $1.72 \times 10^{-8}$ and L2 = $5.33 \times 10^{-8}$.

The study on the wavelength prediction accuracy on the simulated data has been performed for different number of the microresonators changing from 10 to 100. To ensure reliable statistics for assessing the regression model accuracy, microresonators were randomly picked from a set of 200 available cavities to create 30 different subsets for each specified number of microresonators. To guarantee reasonable time constraints for tuning the model, number of repetitions has been restricted to 98 for the entire set of varying splitting levels and resonators number. All 2352 repetitions have been utilized for the prominent datasets to validate the ultimate prediction accuracy. Full statistics on wavelength predictions for simulated spectral datasets is presented in the Fig. 6.

The results for 98 repetitions indicate that the wavelength prediction accuracy is influenced by both the number of microresonators and the eccentricity level (Fig. 6 (a-c)). Specifically, increasing the number of microresonators used for wavelength determination leads to improved precision. This can be clearly seen on example of the microresonators with no mode splitting ($1 \times 10^{-6}$) where the distribution of absolute errors is first dominated by larger error values ($\sim$ 0.1). As the number of microresonators increases, the portion of large error values decreases and the shape of the distribution begins to exhibit two peaks. For instance, with 60 microspheres, one peak is observed around 0.5 and the other around 0.005. When the microresonator spectra exhibit eccentricity-induced splitting, the prediction accuracy improves compared to spectra with sharp peaks. This holds independently on amount of microresonators and becomes more apparent as more microresonators is accounted. For a small number of microresonators, the error distribution may also acquire a more complex shape, with two peaks, as previously described. Regarding the influence of splitting levels on the prediction accuracy, we observed a pattern indicating the existence of an optimal splitting level that yields the highest accuracy with the fewest microresonators accounted. Accuracy reaches a local minimum around an eccentricity level of 0.012, and as the eccentricity level increases further, model performance declines. This agrees perfectly with the strict correlation model provided above (Fig. 5). Notably, for a moderate number of microresonators (up to 60), the 95% quantile for microresonators with significant eccentricity rises by more than one order of magnitude. As the number of microresonators increases, the difference in prediction accuracy across different splitting levels diminishes, and the best performance remains relatively unchanged. Furthermore, the 95% quantile stays at least twice as large (0.6 pm) as the spectral resolution in the dataset. In order to define the ultimate wavelength prediction performance, the whole set of the measurement points has been utilized for the dataset with $e$ = 0.012 for 60 and 80 microresonators (Fig. 6 (d)). In this case, the median prediction accuracy reaches 40 fm, with the 95% quantile at 0.1 pm, demonstrating reliable identification of different wavelengths at the given spectral resolution.

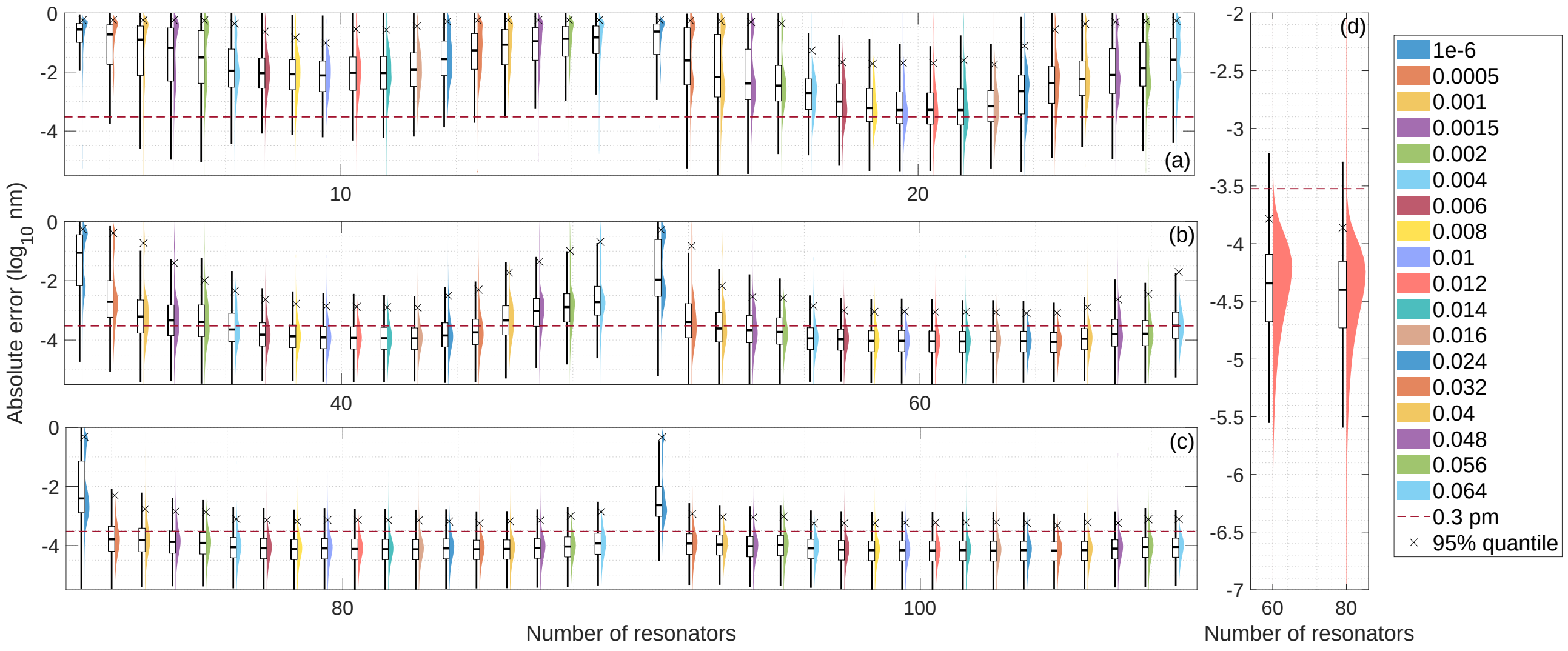

**Fig. 6.** Statistics on the wavelength prediction accuracy in the form of the absolute error value for the simulated data out of 98 measurement points with spectra of different splitting levels ($e \in 1\times10^{-6}, 6.4\times10^{-2}$) and varying number of microresonators in the range of [10:100] that were utilized to build the regression model (a-c). Red horizontal dashed line indicates the spectral resolution of simulations (0.3 pm) and black crosses show the 95% quantile of the corresponding distributions. The prediction accuracy for the spectral data for $e$ = 0.012, which was obtained by accounting 2352 measurement points (d).

In order to experimentally validate the advantages of microcavities with split resonances in comparison to those with sharp peaks, two sets of spectra were collected. They were obtained from the same chip with 5563 microspheres in the unprocessed (original) state and after thermal treatment at 160°C to achieve the eccentricity induced splitting. To ensure comparability with the simulated results and maintain a reasonable time for tuning the wavelength regression model, the prediction range was limited to 1.2 nm, consisting of 2352 measurement repetitions. The full set of all available microresonators was used to generate subsets with varying numbers (0-100) of randomly picked microspheres. In total 30 subsets for each cavities number have been formed to ensure reliable statistics on regression model performance. Since the simulated spectrum involves certain assumptions, the optimal LightBoost parameters for the simulated dataset may not yield the highest prediction accuracy for experimental spectra. Therefore, the model settings were extra optimized with the learning rate set to 0.1, the maximum depth of 30, number of leaves equal to 2560, and regularization parameters L1 and L2 set to $1.14\times10^{-6}$ and $2.16\times10^{-8}$, respectively. Statistics on the accuracy of the wavelength prediction for both experimental sets is shown in Fig. 7.

As anticipated, the results show that increasing the number of microresonators enhances wavelength detection accuracy. Notably, cavities with split resonances outperform those with sharp resonances independently on amount of the cavities considered. Specifically, when a small number of microcavities with sharp resonances are considered, the absolute error distribution contains several peaks. One of them represents the substantial errors at the ~100 pm level and indicates that many wavelengths remain uncertain. For example, when only 10 cavities are used to build the regression model, these errors dominate, but their portion decreases with the number of the microresonators considered. When 100 cavities with sharp peaks are used to train the regression model, particular wavelengths, however, remain challenging to accurately estimate, demonstrating outlines at ∼100 pm level. In contrast, the regression model trained on microresonators with induced splitting do not exhibit this behavior and demonstrates close to the normal distribution in the logarithmic scale independently on microresonator number. Here, the spread in the accuracy of the wavelength identification decreases gradually with the number of microresonators. The regression model trained on 40 microresonators with mode splitting results in the median error value dropping below the wavelength resolution (0.3 pm), whereas the 95% quantile remains one order of magnitude beyond. In this case and for 60 microresonators in the set, the discrepancy in the prediction accuracy between the split and sharp spectra is the most pronounced and reaches two orders of magnitude. By increasing the number of resonators with the split spectra up to 100, the median accuracy reaches the level of 80 fm with 95% quantile at 0.4 pm, and the wavelength prediction outperforms by at least one order of magnitude compared to the case of sharp spectra. With 200 microresonators in the set, the 95% quantile reaches 0.2 pm and thus is below the spectral resolution (not shown in the Fig.7).

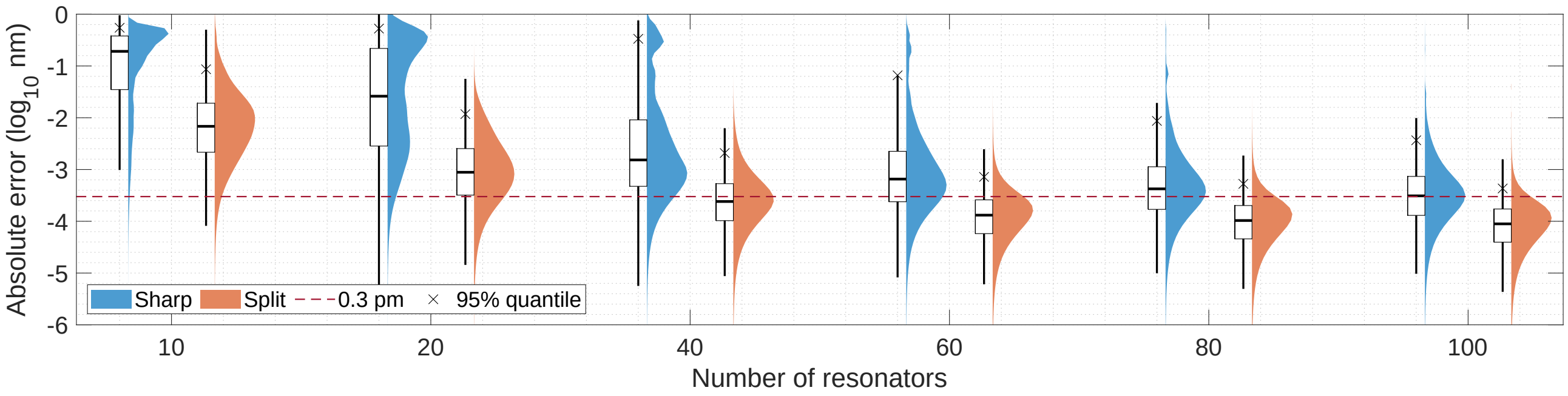

**Fig. 7.** Statistics on the wavelength prediction accuracy, represented as absolute error values, which were obtained from 2352 measurement points for the chip in both its original (sharp) and thermally-treated (split) states, while varying the number of microresonators in the range of [10:100].

The discrepancy in prediction accuracy between simulated (Fig. 6) and experimental (Fig.7) data for the same number of the microresonatros can be attributed to the volatile eccentricity level for microspheres on the sample, which is a consequence of the initial non-spherical geometry of the microcavities. Thus the experimental set of microcavities with sharp peaks demonstrates superior performance for the same number of resonators in comparison to the simulated dataset of almost ideal microspheres. Consequently, experimental microspheres with split resonances, which may exhibit eccentricity values exceeding the optimal value defined for the simulated data, achieve lower prediction accuracy.

## Broadband wavelength prediction

In the data-driven methods, the accuracy and reliability of the predictions strongly depend on the quality, consistency and amount of the training data, where the latter plays a vital role in improving the reliability and generalizability of predictive models. In addition to mitigating the effects of random noise and outliers, having a sufficient set of repetitions enables to learn the underlying patterns more efficiently, which is necessary to achieve high accuracy in complex predictive tasks. To study the effect of the experimental repetitions on wavelength prediction, we constructed several subsets with different numbers of experiment repeats (98, 294, 588, 1176, 2352) out of the total 2352 available. The results of the study for the wavelength prediction accuracy within 1.2 nm spectral region with 200 microresonators and different number of the spectra repetitions available is represented in Fig. 8. The results show that the median error value improves twice down to 50 fm level whereas the 95% quantile level reduced from 680 fm down to 210 fm when accounting 2352 instead of 98 repetitions. Thus, optimal performance for the wavelength prediction is attainable when excessive number of experimental repetitions are available.

In contrast to the wavelength prediction within a 1.2 nm spectral range, where an adequate experimental dataset representing 2532 repetitions can be constructed within hours, expanding the spectral range by already one order of magnitude (∼ 10 nm) necessitates experimental studies lasting many days or even weeks. For this reason, the limited experimental set representing ∼10 nm range has been extended with data augmentation, whereby new responses are generated via VAE. In order to reflect the maximum number of the factors influencing the spectral response and thus to avoid overfitting, sufficiently representative experimental dataset has to be provided. For this, the experimental dataset that covers 10.13 nm spectral range with resolution of 0.3 pm (33765 data points) was measured 41 times consecutively for the same chip with 5563 microspheres. The required number of microresonators is expected to exceed 100 determined for the narrow wavelength prediction to guarantee comparable wavelength prediction accuracy. As the number of features grows, the computational complexity increases exponentially. In order to cover the responses of thousands of microresonators by keeping the data volume moderate, the dataset has been preprocessed with principal component analysis (PCA). It performs dimensionality reduction by transforming the original data into a new set of features (principal components, PC) that capture the most variance. By accounting smaller number of PCs than the number of the microresonators is available, PCA minimizes the input space while preserving the core structure of the data. This allows the prediction model to focus on the most significant information and exclude noise at most. In order to define the PCA transformation matrix, the averaged spectral responses over 41 repetitions have been calculated for all 5563 microspheres available. The transformation matrix is fixed for the chip and is utilized for transition from microresonator space to the principal component space and vice versa for all spectra repetitions. The results of the spectral data preprocessing are represented in Fig. 9.

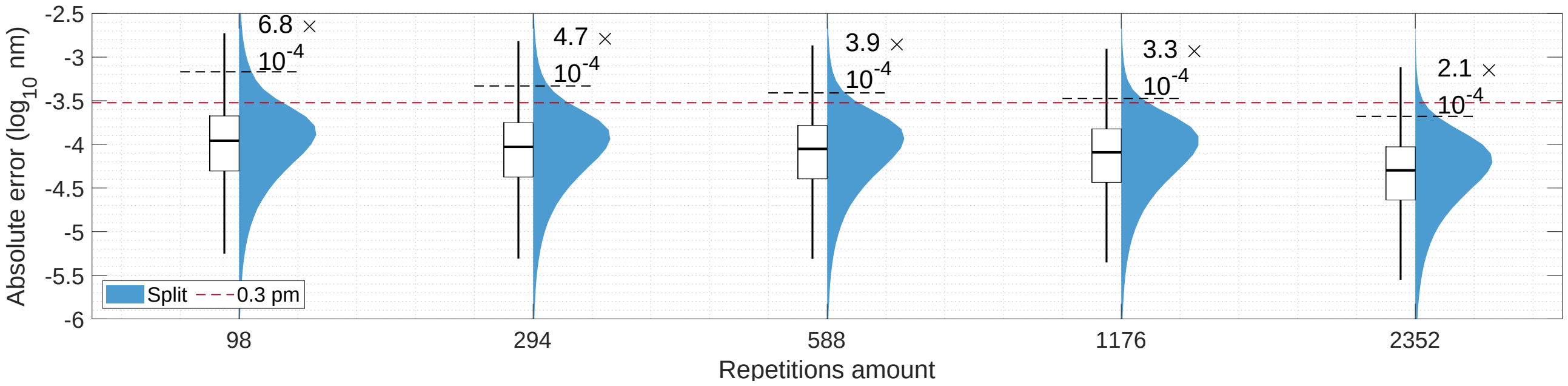

**Fig. 8.** Influence of the number of experimental repetitions on the wavelength prediction accuracy within a narrow spectral band for 200 microresonators.

The averaged spectral response in the space of the two first PCs that accounts in total 19.8% variance in the original data is represented in Fig. 9 a. It has been determined that despite the spectral periodicity of each microresonator and high consistency between periods, already two first PCs show only several crossings across the whole spectral band. By increasing number of the PCs, more spectral data variability becomes reflected in the transformed space. It was shown that with the first 10 PCs, 50% of variability in the original data is represented, with 80 PCs - 90%, with 150 PCs - 95% and with 630 PCs - 99 (Fig. 9 b). We determined that already the first 80 principal components are sufficient to describe effectively the spectral data of the chip with 5563 microspheres. By increasing the number of considered PCs, a moderate portion of the significant information from the initial feature set is added, while leading to a significant complication of the procedure of adaptation of the prediction model. As a result, the original dimensionality of the experimental dataset $41 \times 5563 \times 33765$ is reduced to $41 \times 80 \times 33765$.

To account for the decreasing significance of PCs with their index, we constructed individual VAEs for most significant PCs (from PC1 to PC10), where each reflects more than 2% of the overall variance and in total 50%. The remaining 70 PCs (overall 40%) have been utilized to train the single VAE, where the dataset was reshaped by stacking various PCs. For each of 11 VAEs, 40 repetitions were attributed to the training set, where the first 10 VAEs are trained using a batch size of 8. The 11th VAE is trained with a batch size of 28, for each of which samples are randomly selected and undergoing iterations for weight updates during each epoch. This guarantees that all samples in the dataset were processed within each epoch and each batch contains samples from different PCs, thereby enabling $VAE11$ to learn diverse PC patterns. Given the limited number of repetitions in the experimental data set, it was not feasible to divide the data into training and validation parts. Allocating at least 20% to validation would have substantially reduced the effectiveness of the first 10 VAEs. For this reason, we validated the models with 50 VAE outputs generated by using averaged experimental response as the input. Thus, in addition to monitoring the total loss derived as MSE and KLD, we also monitored MAE between the input and the reconstructed data. As a result, 11 trained VAEs have been utilized to augment 41 experimental measurements up to 2352 repetitions to be employed for the regression model training.

The broadband wavelength prediction was performed using the same architecture as previously defined for the narrowband data. Statistics on the wavelength prediction accuracy within the 10 nm broad range utilizing VAE augmented dataset is represented in Fig. 10. In this study, we evaluated the wavelength estimation performance by comparing the results obtained utilizing the first 10 and 80 PCs, that represent 50% and 90% of the overall experimental variability, respectively. The results show that by feeding 10 features to the regression model the median value for prediction accuracy reaches 100 fm with the 95% quantile at 0.6 pm. Considering 80 PC, the median prediction accuracy improves insignificantly, but each wavelength in the 10 nm spectral range can be unambiguously identified with a certain spectral resolution of 0.3 pm. Thus, the accuracy gain from 10 to 80 PCs is less pronounced than previously shown between 10 and 80 resonators within 1.2 nm spectral window (Fig. 7). This opens up the prospect of reduction of the PCs number down to the first 10 which, in turn, allows for extension of the wavelength prediction region up to ~100 nm by keeping the computational efforts to build the regression model at the same level. The average time required for the pre-trained regression model to recover a specific wavelength with known 80 PC values is 80 $\mu$s.

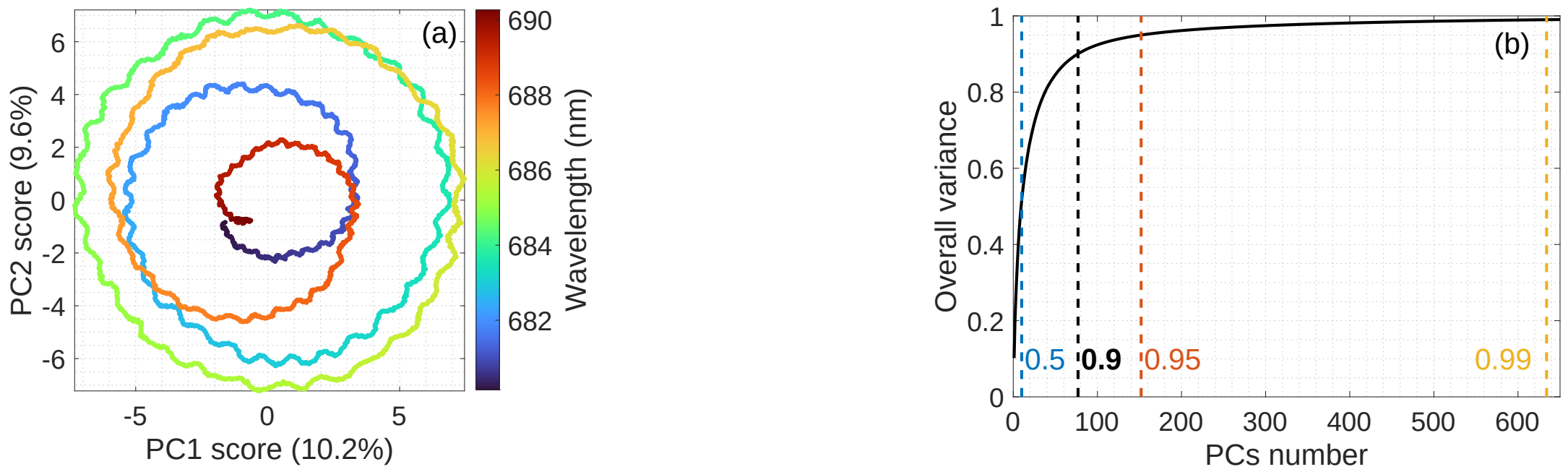

**Fig. 9.** Results of preprocessing of the 10 nm wide spectral data with principal component (PC) analysis. (a) Averaged spectral response in the space of the first two PCs, where the color spectrum reflects the wavelength changes in the examined band. (b) The dependence of the represented variance in the original data depending on the number of the PCs accounted (vertical lines represent 50, 90, 95 and 99% levels).

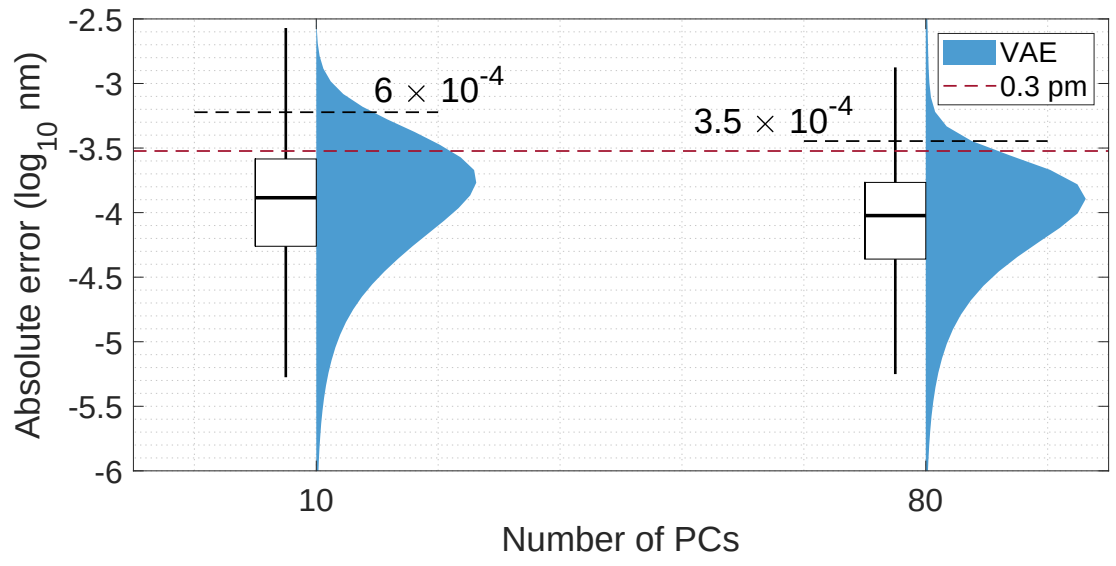

**Fig. 10.** Statistics on the wavelength prediction accuracy for the 10 nm spectral range with VAE augmented experimental dataset comprising 2352 repetitions with the varied number of the features (10, 80).

## Discussion

In this paper, we proposed a novel type of the high-precision compact and affordable reconstructive wavemeter that is based on the induction of the resonance mode splitting by controllable deforming of the spherical microcavities for the high-resolution prediction. The proposed instrument is based on integration of thousands of the PMMA microspheres on a single substrate for simultaneous response collection. We proposed to utilize eccentricity mode splitting of the whispering gallery modes which is enabled by thermal processing of the chip to establish a unique wavelength-dependent pattern for the broad wavelength range. We have studied the observational probabilities of the mode splitting phenomenon for raw microresonators and developed a mechanism for multiplicative growth of the fraction of microresonators on a chip with mode splitting from $\approx$ 1% up to $\approx$ 45%. We have demonstrated the existence of an optimal level of microsphere deformation (1.2%) to attain the ultimate wavemeter performance. The intrinsic sensitivity for the deformed microresonators can be improved by at least one order of magnitude than for the chips with ideal microspheres and approaches 300 fm. We have demonstrated that a set of only 100 mode-splitting resonators is sufficient to attain this accuracy level in wavelength prediction within the range of ~nm. Ultimately, the proposed instrument was validated for the broadband wavelength estimation through the implementation of the spectral response densification via the dimensionality reduction mechanism, data augmentation with a variation autoencoder, and the ensemble boosting method as a regression model. This approach ensured the wavelength determination at the ~100 fm level within the ~10 nm spectral range, with the potential for its expansion up to ~100 nm.

With respect to the intrinsic sensitivity, which is independent of the interpretation technique, the performance of the reconstructing wavemeter based on a disordered set of deformed microresonators is approaching the ultimate performance demonstrated thus far for an integration sphere used as a source for wavelength-dependent speckle formation. However, unlike the integrating spheres behind physical principle of reconstructing wavemeter, the disordered set of deformed microspheres is superior for miniaturization and enables chip-scale integrability. It is also important to note that similarly to the other configurations such as based on multimode fiber [48] or integrating sphere [22] the precision limit can be lowered more than six orders of magnitude below the intrinsic sensitivity level by extracting latent variables from the raw speckles within the limited spectral window. Being already introduced, principal component analysis as a method to enable construction of the representative dataset, the same method can be applied to diminish the correlation between closely spaced wavelengths. Therefore, the precision limit for the proposed approach can be moved toward sub-fm level by using a sophisticated configuration of the calibration light source. In particular, the laser locked to the Rubidium line and supplemented by acousto-optic modulator for tuning the laser radiation can provide a controlled wavelength shift within $\sim$ 10 fm and with minimal steps at ultimate several attometer level.

## Materials and methods

### Multiplexed microresonator chip

As a single detection unit of the multiplexed microresonator chip we have studied commercially available soda-lime glass (Cospheric LLC) and polymethyl methacrylate, PMMA (MicroParticles GmbH) microspheres. Dimensions of the microspheres in the supplied batch are between 50 and 120 $\mu$m. The microspheres undergo cleaning procedure to minimize light scattering during resonance excitation and then are allocated on the 150 $\mu$m glass substrates within an area of 10$\times$10 mm. Relative dimensions of the microresonators and the monitoring area allows allocation of up to several thousands of microcavities. The substrate is chosen to match the optical prism for optical consistency and is carefully cleaned before microcavity deposition. In order to ensure the long-term signal stability, the microspheres are bonded with the substrate via low refractive index films out of the MY-133MC (MyPolymer) material. Low refractive index material is deposited onto the substrate and spin-coated to form a uniform film. Until it remains soft, the microspheres are placed by free fall and become partially embed into the film. Moderate thickness and uniformity of the fixation layer has been identified as a pivotal step in the chip fabrication, where insufficient layer thickness may compromise the stability of the WGM signal and excessive thickness can degrade the spectral properties of the cavities. A film thickness of 2 $\mu$m is deemed the minimum necessary to guarantee the robustness of the microresonators-substrate bond. In order to minimize the coupling loss for the microcavities and thus maximize the loaded Q-factor, a separate spacer layer is required to keep the microspheres away from the substrate. To maintain consistent distancing between the microresonators and the substrate within several hundred nanometers, the original water-matched solution was strongly diluted with the solvent. It has been determined that the spacer layer of $\approx$400 nm enables to enhance the loaded Q-factor from $\sim 10^4$ observable for non-processed substrates up to $\sim 10^6$. Finally, the substrate with immobilized microcavities is covered by the layer of low refractive index fluoropolymer CYTOP (AGC Inc.) with refractive index of 1.34. CYTOP layer was deposited by multiple spin-coating steps on the substrates to dip microspheres and served several functions. First, it is dedicated to avoid contamination of the microspheres and consequently the long-term degradation of their spectral properties, secondly to harmonize the refractive index with the fixation layer and thus guarantee effectively a single medium around the microresonators. An example of the multiplexed microresonator chip with more than 5000 PMMA microspheres is shown in Fig. 11. The chip is characterized by varying cavity sizes and their random arrangement.

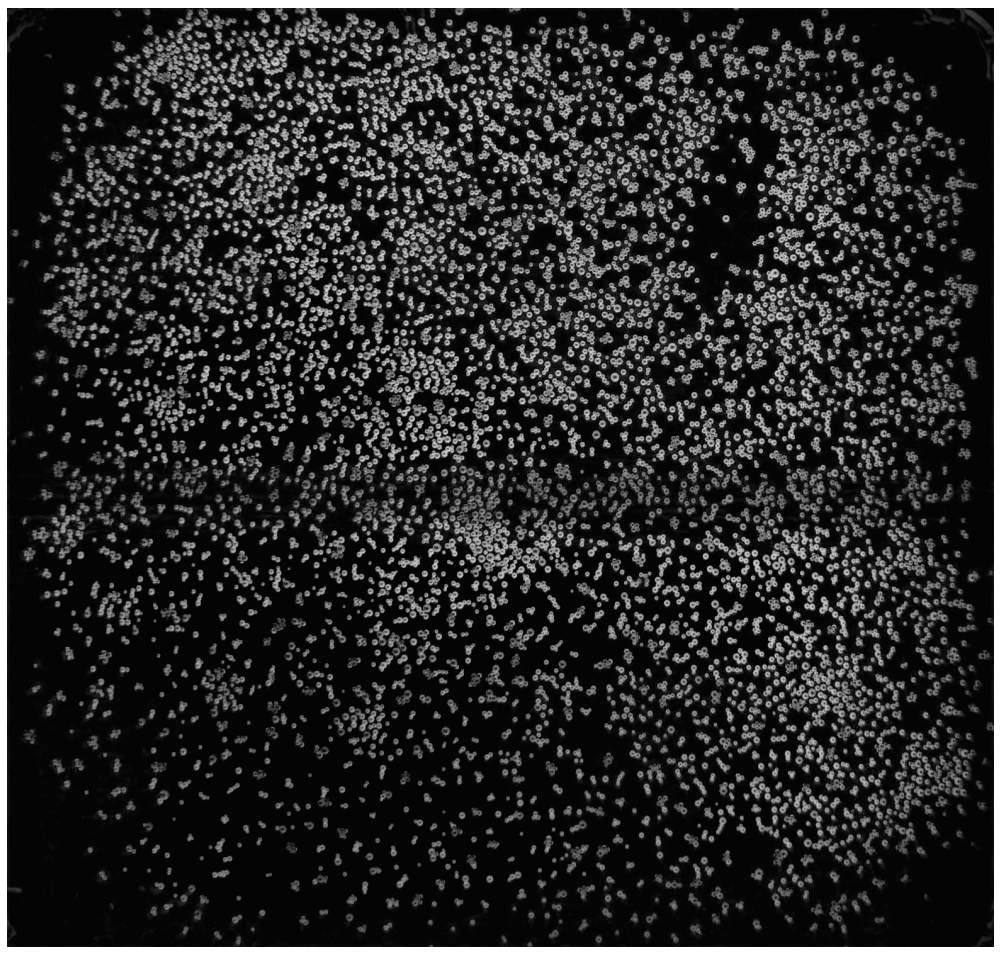

**Fig. 11.** Overview of the multiplexed microresonator chip with PMMA microspheres. Demonstrated image is captured by the measuring instrument and is processed for denoising and background glow removal procedures to enhance clarity (as described in [49]).

### Instrument

The collection of WGM signals from a measuring chip is accomplished using a previously established multiplexed microresonator imaging method [36, 37, 39]. In this approach, resonances in all cavities are simultaneously excited via single optical prism (see Fig. 12). The measuring chip with embedded microspheres is placed onto the optical prism using immersion oil to maintain stable coupling conditions. Instead of conventional photodiode-based monitoring of the excitation channel, the radiated signal from each individual microresonator is captured. A monochrome high-speed global shutter camera (CB262RG-GP-X8G3, Ximea) equipped with lens objective enables signal collection along with spatial distinction of signals from individual microspheres. To equalize the excitation conditions across the microspheres, the chip is illuminated within a collimated beam confined to $\approx$ 8 mm, as determined by the achromatic optical collimation package (60FC-T-4-M40-24, Schaefter+Kirchhoff). The elongated collimated beam along the propagation direction at the prism excitation surface is compensated by transforming the laser beam profile from circular to elliptical. To minimize the impact of thermal fluctuations prism chip assembly is thermostabilized with TEC controller (TEC-1091, Meerstetter). Calibration measurements are performed with tunable diode laser (Velocity, New Focus) within [680:690] nm range. The beam profile is shaped by passing through a corresponding single-mode fiber. Beam polarization is adjusted using a controller (FPC030, Thorlabs) to excite TE modes in the microcavities. Fizeau configuration-based wavemeter with an absolute accuracy of 30 MHz (WS7-30, HighFinesse) has been utilized as a reference unit.

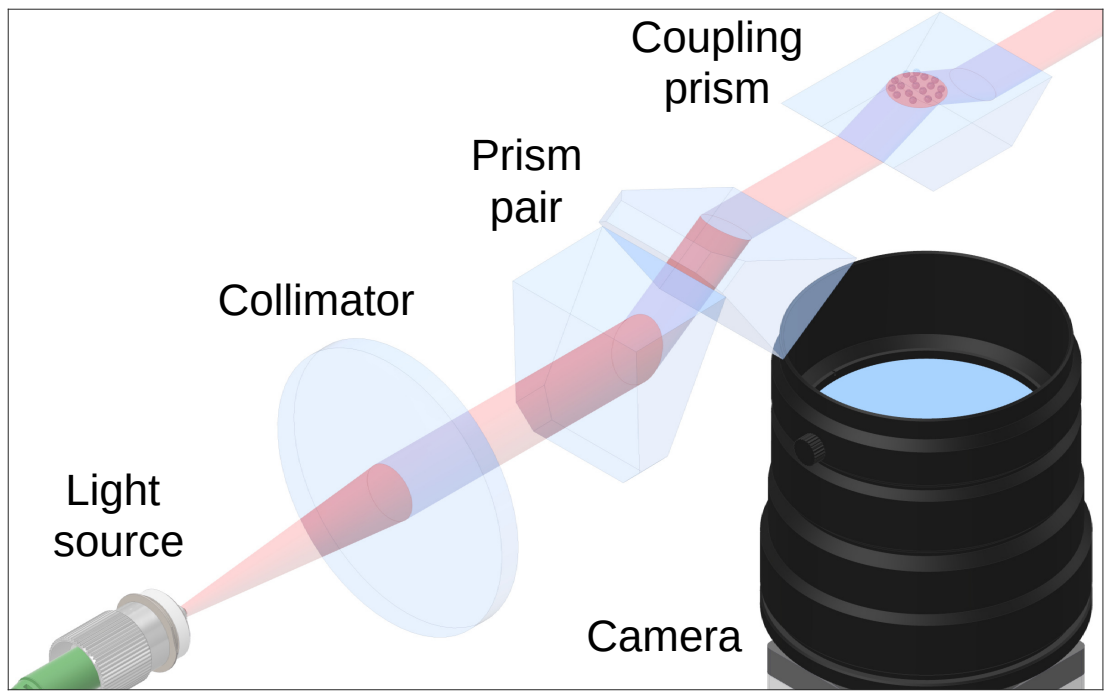

**Fig. 12.** Multiplexed microresonator imaging method for wavemeter with eccentricity-splitted microspheres.

### Data collection

In order to match the variations of the radiated intensities with the particular microresonators on the chip, a map of the microcavities is established. This is regarded as a computer vision task where numerous circular objects are identified in the image, as we outlined in [49]. The process consists of a pipeline with multiple preprocessing steps followed by the circular Hough Transform. The preprocessing steps involve image denoising using the BM3D method [50], image sharpening to enhance edge visibility, background correction, and histogram equalization. As a result, a clear image with high contrast between the background and the microresonators is produced (see Fig. 11), which is then used as the input for a circular Hough Transform. An optimized run achieves the detection accuracy of over 95%, with a slight decrease in accuracy attributed to the Hough Transform's tendency to underestimate. These instances have a negligible impact on the data representativity.

The radiated light integrated over the whole microsphere area is considered as the microcavity signal. Its strength is defined by location of the laser frequency on the WGM spectrum which is unique among the cavities on the chip due to natural variation in their properties. As a result, each frequency of the laser illuminating the measuring chip is matched to the set of intensities with dimensionality corresponding to the number of resonators. The experimental dataset has been constructed by tuning the laser frequency over the predefined range multiple times with simultaneous tracking of the intensity distribution over the chip. The laser tuning speed and the camera frame rate has been adjusted to ensure both reasonable aquisition time to get a single spectrum and spectral resolution $\sim$100 fm which is at least twice as much as the absolute accuracy of the calibration wavemeter. Finally, to harmonize the spectral data between the measurements, they are aligned to a common wavelength grid with a step of 300 fm.

### Acknowledgement

The authors Andreas Ostendorf and Anton Saetchnikov are grateful to the German Federal Ministry for Research and Education (BMBF) for partly funding this work under the VIP+-Programme in the project IntellOSS, 03VP08220.

### Data availability

The data that support the findings of this study are available from the corresponding author upon reasonable request.

### Conflict of interest

The authors declare no conflicts of interest.

## References

[1] Udem, T., Holzwarth, R. & Hänsch, T. W. Optical frequency metrology. *Nature* **416**, 233–237 (2002).

[2] Couturier, L. et al. Laser frequency stabilization using a commercial wavelength meter. *The Review of scientific instruments* **89**, 043103 (2018).

[3] Newman, Z. L. et al. Architecture for the photonic integration of an optical atomic clock. *Optica* **6**, 680 (2019).

[4] de Leon, N. P. et al. Materials challenges and opportunities for quantum computing hardware. *Science* **372** (2021).

[5] Jørgensen, A. A. et al. Petabit-per-second data transmission using a chip-scale microcomb ring resonator source. *Nature Photonics* **16**, 798–802 (2022).

[6] Dobosz, M. & Kożuchowski, M. Overview of the laser-wavelength measurement methods. *Optics and Lasers in Engineering* **98**, 107–117 (2017).

[7] Coluccelli, N. et al. The optical frequency comb fibre spectrometer. *Nature communications* **7**, 12995 (2016).

[8] Niu, R. et al. khz-precision wavemeter based on reconfigurable microsoliton. *Nature communications* **14**, 169 (2023).

[9] Xu, B. et al. Whispering-gallery-mode barcode-based broadband sub-femtometer-resolution spectroscopy with an electro-optic frequency comb. *Advanced Photonics* **6**, SM1D–4 (2024).

[10] Gardner, J. L. Compact fizeau wavemeter. *Applied optics* **24**, 3570 (1985).

[11] Rotter, S. & Gigan, S. Light fields in complex media: Mesoscopic scattering meets wave control. *Reviews of Modern Physics* **89**, 648 (2017).

[12] Yang, Z. et al. Miniaturization of optical spectrometers. *Science* **371** (2021).

[13] Wan, N. H. et al. High-resolution optical spectroscopy using multimode interference in a compact tapered fibre. *Nature communications* **6**, 7762 (2015).

[14] Bruce, G. D. et al. Femtometer-resolved simultaneous measurement of multiple laser wavelengths in a speckle wavemeter. *Optics Letters* **45**, 1926–1929 (2020).

[15] Gao, Z. et al. Breaking the speed limitation of wavemeter through spectra-space-time mapping. *Light: Advanced Manufacturing* **4**, 1 (2024).

[16] Wang, T. et al. Decoding wavelengths from compressed speckle patterns with deep learning. *Optics and Lasers in Engineering* **180**, 108268 (2024).

[17] Redding, B. et al. Compact spectrometer based on a disordered photonic chip. *Nature Photonics* **7**, 746–751 (2013).

[18] Wang, Z. et al. Single-shot on-chip spectral sensors based on photonic crystal slabs. *Nature communications* **10**, 1020 (2019).

[19] Brown, C. et al. Neural network-based on-chip spectroscopy using a scalable plasmonic encoder. *ACS nano* **15**, 6305–6315 (2021).

[20] Sun, Q. et al. Compact nano-void spectrometer based on a stable engineered scattering system. *Photonics Research* **10**, 2328 (2022).

[21] Metzger, N. K. et al. Harnessing speckle for a sub-femtometre resolved broadband wavemeter and laser stabilization. *Nature communications* **8**, 15610 (2017).

[22] Gupta, R. K. et al. Deep learning enabled laser speckle wavemeter with a high dynamic range. *Laser & Photonics Reviews* **14**, 2579 (2020).

[23] Popoff, S. M. et al. Measuring the transmission matrix in optics: An approach to the study and control of light propagation in disordered media. *Physical Review Letters* **104**, 100601 (2010).

[24] Kwak, Y. et al. A pearl spectrometer. *Nano letters* **21**, 921–930 (2021).

[25] Edrei, E. et al. Chip-scale atomic wave-meter enabled by machine learning. *Science advances* **8**, eabn3391 (2022).

[26] Facchin, M. et al. Determining intrinsic sensitivity and the role of multiple scattering in speckle metrology. *Nature Reviews Physics* **6**, 500–508 (2024).

[27] Braginsky, V. B., Gorodetsky, M. L. & Ilchenko, V. S. Quality-factor and nonlinear properties of optical whispering-gallery modes. *Physics Letters A* **137**, 393–397 (1989).

[28] Vahala, K. J. Optical microcavities. *Nature* **424**, 839–846 (2003).

[29] Loyez, M. et al. From whispering gallery mode resonators to biochemical sensors. *ACS sensors* **8**, 2440–2470 (2023).

[30] Tang, S.-J. et al. Single-particle photoacoustic vibrational spectroscopy using optical microresonators. *Nature Photonics* **17**, 951–956 (2023).

[31] Zossimova, E. et al. Whispering gallery mode sensing through the lens of quantum optics, artificial intelligence, and nanoscale catalysis. *Applied Physics Letters* **125**, 13832 (2024).

[32] Wan, Y. et al. Reconstructive spectrum analyzer with high–resolution and large–bandwidth using physical–model and data–driven model combined neural network. *Laser & Photonics Reviews* **17** (2023).

[33] Schweiger, G., Nett, R. & Weigel, T. Microresonator array for high-resolution spectroscopy. *Optics Letters* **32**, 2644–2646 (2007).

[34] Petermann, A. B. et al. Surface-immobilized whispering gallery mode resonator spheres for optical sensing. *Sensors and Actuators A: Physical* **252**, 82–88 (2016).

[35] Berkis, R. et al. Wavelength sensing based on whispering gallery mode mapping. *Fibers* **10**, 90 (2022).

[36] Saetchnikov, A. V. et al. Reusable dispersed resonators-based biochemical sensor for parallel probing. *IEEE Sensors Journal* **19**, 7644–7651 (2019).

[37] Saetchnikov, A. et al. Deep-learning powered whispering gallery mode sensor based on multiplexed imaging at fixed frequency. *Opto-Electronic Advances* **3**, 200048 (2020).

[38] Saetchnikov, A. V. et al. Intelligent optical microresonator imaging sensor for early stage classification of dynamical variations. *Advanced Photonics Research* **7**, 2100242 (2021).

[39] Saetchnikov, A. V. et al. Detection of per- and polyfluoroalkyl water contaminants with a multiplexed 4d microcavities sensor. *Photonics Research* **11**, A88 (2023).

[40] Sumetsky, M. Whispering-gallery-bottle microcavities: The three-dimensional etalon. *Optics Letters* **29**, 8–10 (2004).

[41] Gorodetsky, M. L. & Fomin, A. E. Geometrical theory of whispering-gallery modes. *IEEE Journal of Selected Topics in Quantum Electronics* **12**, 33–39 (2006).

[42] Ilchenko, V. S. et al. Whispering gallery mode diamond resonator. *Optics Letters* **38**, 4320–4323 (2013).

[43] Xie, Y. et al. Batch fabrication of high-quality infrared chalcogenide microsphere resonators. *Small* **17**, e2100140 (2021).

[44] Le Thomas, N. et al. Effect of a dielectric substrate on whispering-gallery-mode sensors. *Journal of the Optical Society of America B* **23**, 2361–2365 (2006).

[45] Francois, A. & Himmelhaus, M. Optical sensors based on whispering gallery modes in fluorescent microbeads: size dependence and influence of substrate. *Sensors (Basel, Switzerland)* **9**, 6836–6852 (2009).

[46] Gorodetsky, M. L. & Ilchenko, V. S. Optical microsphere resonators: Optimal coupling to high-q whispering-gallery modes. *Journal of the Optical Society of America B* **16**, 147 (1999).

[47] Gorodetsky, M. L., Pryamikov, A. D. & Ilchenko, V. S. Rayleigh scattering in high-q microspheres. *Journal of the Optical Society of America B* **17**, 1051–1057 (2000).

[48] Bruce, G. D. et al. Overcoming the speckle correlation limit to achieve a fiber wavemeter with attometer resolution. *Optics Letters* **44**, 1367–1370 (2019).

[49] Saetchnikov, A. et al. Mapping of the detecting units of the resonator-based multiplexed sensor. In Optical Micro- and Nanometrology VII, Vol. 10678, 106780W (SPIE, 2018).

[50] Dabov, K. et al. Bm3d image denoising with shape-adaptive principal component analysis. *Proc. Workshop on Signal Processing with Adaptive Sparse Structured Representations (SPARS'09)* (04 2009).